\documentclass{article}
\usepackage[preprint]{tmlr}
\usepackage{graphicx}
\usepackage{subcaption}
\usepackage[T1]{fontenc}    % use 8-bit T1 fonts
\usepackage{hyperref}       % hyperlinks
\usepackage{url}            % simple URL typesetting
\usepackage{booktabs}       % professional-quality tables
\usepackage{amsfonts}       % blackboard math symbols
\usepackage{nicefrac}       % compact symbols for 1/2, etc.
\usepackage{microtype}      % microtypography
\usepackage{xcolor}         % colors
\usepackage{enumitem}

\hypersetup{colorlinks,linkcolor={red},citecolor={blue}}  

\usepackage{tabularx}
\usepackage{multirow}
\usepackage{array}
\usepackage{makecell}

\title{
\begin{center}
BIASeD: Bringing Irrationality into Automated System Design
\end{center}}

\author{\name Aditya Gulati \email aditya@ellisalicante.org \\
        \addr ELLIS Alicante
        \ANDAUTH
        \name Miguel Angel Lozano \email malozano@ua.es \\
        \addr Universidad de Alicante
        \ANDAUTH
        \name Bruno Lepri \email lepri@fbk.eu \\
        \addr Fondazione Bruno Kessler
        \ANDAUTH
        \name Nuria Oliver \email nuria@ellisalicante.org \\
        \addr ELLIS Alicante
}

\begin{document}

\maketitle

\begin{abstract}
Human perception, memory and decision-making are impacted by tens of cognitive biases and heuristics that influence our actions and decisions. Despite the pervasiveness of such biases, they are generally not leveraged by today's Artificial Intelligence (AI) systems that model human behavior and interact with humans. In this theoretical paper, we claim that the future of human-machine collaboration will entail the development of AI systems that model, understand and possibly replicate human cognitive biases. We propose the need for a research agenda on the interplay between human cognitive biases and Artificial Intelligence. We categorize existing cognitive biases from the perspective of AI systems, identify three broad areas of interest and outline research directions for the design of AI systems that have a better understanding of our own biases.
\end{abstract}

\maketitle

\section{Introduction}

A cognitive bias is a systematic pattern of deviation from rationality that occurs when we process, interpret or recall information from the world, and it affects the decisions and judgments we make. Cognitive biases may lead to inaccurate judgments, illogical interpretations and perceptual distortions. Thus, they are also referred to as \emph{irrational} behavior \cite{Kahneman1979,Ariely2008}. 

Since the 1970s, scholars in social psychology, cognitive science, and behavioral economics have carried out studies aimed at uncovering and understanding these apparently irrational elements in human decision making. As a result, different theories have been proposed to explain the source of our cognitive biases.

In 1955, Simon proposed the theory of \emph{bounded rationality} \cite{Simon1955}. It posits that human decision making is rational, but limited by our computation abilities which results in sub-optimal decisions because we are unable to accurately solve the utility function of all the options available at all times. Alternative theories include the \emph{dual process theory} and the \emph{prospect theory}, both proposed by Kahneman \cite{Kahneman2011,Kahneman1979}.

Even though there is no unified theory of our cognitive biases, it is clear that we use multiple shortcuts or heuristics\footnote{While heuristics typically refer to a simplifying rule used to make a decision and a cognitive bias refers to a consistent pattern of deviation in behavior, in this paper both terms are used  interchangeably as both impact human decisions in a similar way.} to make decisions which might lead to sub-optimal outcomes. However and despite these limitations, cognitive biases and heuristics are a crucial part of our decision making. 

In fact, cognitive biases have traditionally been commercially leveraged in different sectors to manipulate human behavior. Examples include casinos \cite{Schull2012}, addictive apps \cite{Eyal2014}, advertisement and marketing strategies to drive consumption \cite{Petticrew2020,Ariely2008} and social media campaigns to impact the outcome of elections \cite{Epstein2015}. However, we advocate in this paper for a constructive and positive use of cognitive biases in technology, moving from manipulation to collaboration. We propose that considering our cognitive biases in AI systems could lead to more efficient human-AI collaboration. 

Nonetheless, there has been limited research to date on the interaction between human biases and AI systems, as recently highlighted by several authors~\cite{Hiatt2017,natural_stupidity,Rastogi2022,Kliegr2021}. In this context, we highlight the work by Akata et al.~\cite{Akata2020} who propose a research agenda for the design of AI systems that collaborate with humans, going beyond a human-in-the-loop setting. They pose a set of research questions related to how to design AI systems that collaborate with and adapt to humans in a responsible and explainable way. In their work, they note the importance of understanding humans and leveraging AI to mitigate biases in human decisions.

In this paper, we build from previous work by proposing a taxonomy of cognitive biases that is tailored to the design of AI systems. Furthermore, we identify a subset of 20 cognitive biases that are suitable to be considered in the development of AI systems and outline three directions of research to design cognitive bias-aware AI systems.

\section{A Taxonomy of Cognitive Biases}
\label{sec:biases}

Since the early studies in the 1950s, approximately 200 cognitive biases have been identified and classified \cite{benson_codex,Hubbard2022}. Several taxonomies of cognitive biases have been proposed in the literature, particularly in specific domains, such as medical decision making \cite{BlumenthalBarby2014,Saposnik2016}, tourism~\cite{Wattanacharoensil2019} or fire evacuation~\cite{Kinsey2018}.
Alternative taxonomies classify biases based on their underlying phenomenon  \cite{Tversky1974,Stanovich2008,Arnott2006}. However, given that there is no widely accepted theory of the source of cognitive biases \cite{Pohl2004}, classifying them according to their hypothesized source might be misleading. 

\begin{figure}
\centering
\includegraphics[width=\linewidth]{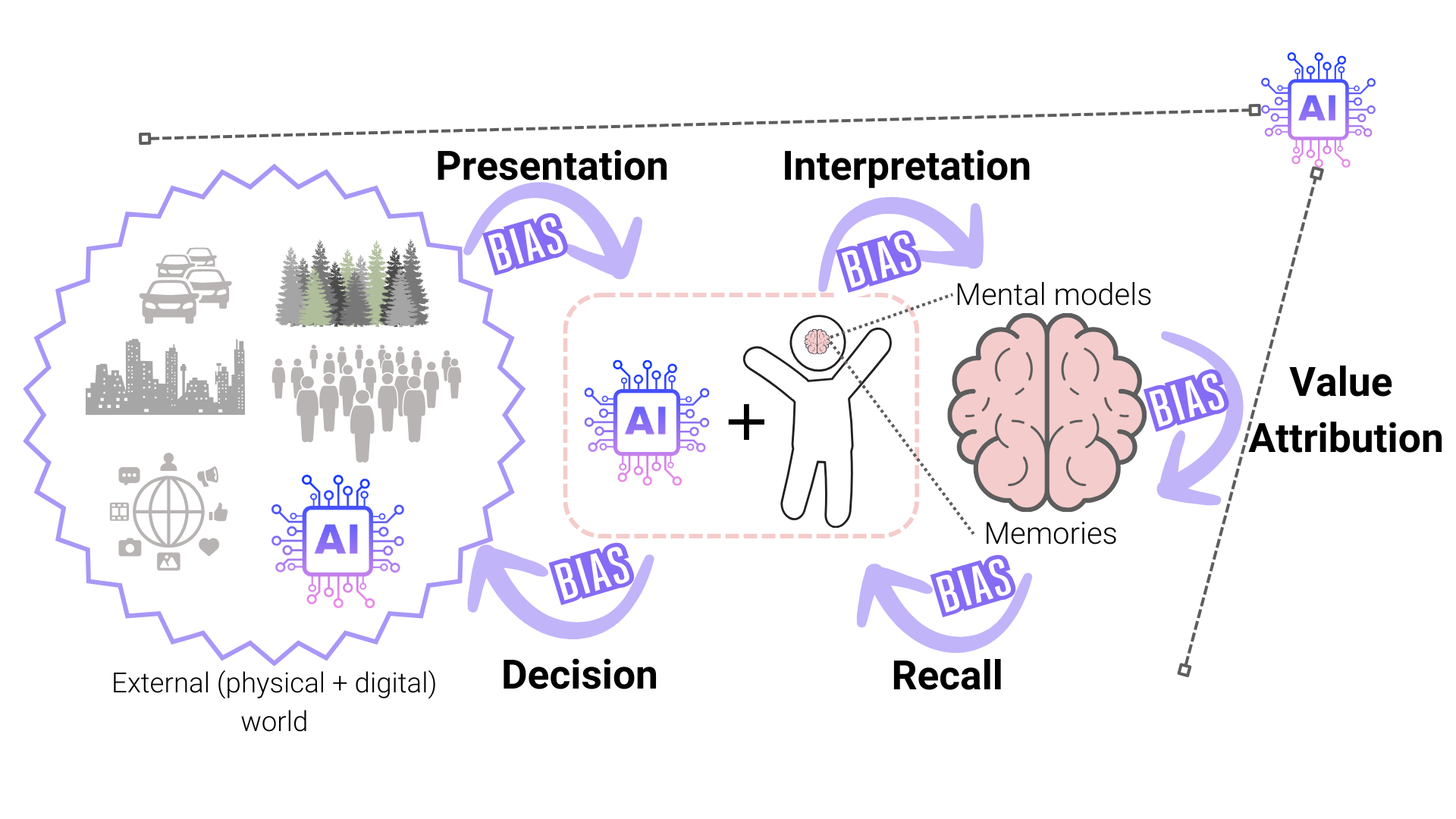}
\caption{Stages of the human perception, interpretation and decision-making process that are impacted by cognitive biases. AI systems (represented by an orange undirected graph) could observe our behavior, detect biases and help us mitigate them.}
\label{fig:biasesTaxonomy}
\end{figure}

Dimara et al.~\cite{viz_taxonomy} report similar limitations with existing taxonomies and propose a new taxonomy of cognitive biases based on the experimental setting where each bias was studied and with a focus on visualization. While this taxonomy is of great value for visualization, our focus is the interplay between AI and cognitive biases. Thus, we propose classifying biases according to five stages in the human decision making cycle as depicted in Figure \ref{fig:biasesTaxonomy}. 

The left part of Figure \ref{fig:biasesTaxonomy} represents the physical world that we perceive, interpret and interact with. The right part represents the internal models and memories that we create based on our experience. As seen in Figure \ref{fig:biasesTaxonomy}, we propose classifying biases according to five main stages in the human perception, interpretation and decision making process:
\emph{presentation} biases, associated with how information or facts are presented to humans; \emph{interpretation} biases that arise due to misinterpretations of information; \emph{value attribution} biases that emerge when humans assign values to objects or ideas that are not rational or based on an underlying factual reality; \emph{recall} biases associated with how we recall facts from our memory and \emph{decision} biases that have been documented in the context of human-decision making.

Figure \ref{fig:biasesTaxonomy} also illustrates how AI systems (represented as an orange undirected graph) may interact with humans in this context. First, AI systems could  be entities in the external world that humans perceive or interact with (e.g. chatbots, robots, apps...). Second, they may be active participants and assist humans in their information processing and decision-making processes (e.g. cognitive assistants, assistive technologies...). Finally, AI systems could be observers that model our behavior and provide feedback without directly being involved in the decision making process. Note that these three forms of interaction with AI systems may occur simultaneously. 

We also present four representative cognitive biases for each category. These biases were chosen according to the amount of evidence in the literature about the existence of the bias and their relevance for the design of AI systems. Tables \ref{table:biasesTablePart1}, \ref{table:biasesTablePart2} and \ref{table:biasesTablePart3} summarize the selected biases, their description, supporting literature and relevance to AI. Additionally, table \ref{tab:conf} illustrates how AI could potentially provide support in detecting and mitigating some of these biases using the confirmation bias as an example.

\begin{table}[t]
\centering
\resizebox{\linewidth}{!}{

\begin{tabular}{ l|p{0.13\linewidth}|p{0.37\linewidth}|p{0.50\linewidth}}

\toprule

 & \multicolumn{1}{c|}{\small\textbf{Bias}} & \multicolumn{1}{c|}{\small\textbf{Brief Description}} & \multicolumn{1}{c}{\small\textbf{Relevance to AI}}\\

\midrule

%------------------------(Presentation)------------------------

\parbox[t]{2mm}{\multirow{4}[45]{*}{\rotatebox[origin=c]{90}{\small\textbf{Presentation}}}}

& \emph{Decoy effect} \cite{Huber1982,Hu2014,Wang2018,Josiam1995} & Placing deliberately a worse alternative between two choices can reverse the user's preference & Could AI systems learn to place decoys effectively while presenting alternatives? Could AI systems learn to identify decoys? \cite{Teppan2012}\\ \cmidrule{2-4}
 
& \emph{Framing effect} \cite{Tversky1981,Gchter2009,Levin1998,Gong2013} & How a statement is framed can alter its perceived value & Studies have shown that when humans are placed in human-AI teams, their decisions \cite{Souza2016} and trust \cite{Kim2022} are impacted by the framing effect. Could AI systems learn to frame explanations to make them more trustworthy? \\ \cmidrule{2-4}
 
& \emph{Anchoring effect} \cite{Tversky1974,Ni2019,Yasseri2022} & Human decision making is influenced by certain reference points or anchors & The use of anchors to alter user preferences has been studied in marketing and recommender systems \cite{Adomavicius2013}. Could AI systems automatically identify anchors that humans might be subject to?  \\ \cmidrule{2-4}
 
& \emph{Pseudocertainty effect} \cite{Tversky1981,Hayes2009,Burakov2013} & Humans incorrectly estimate the certainty of statements in a multi-stage decision making process & Could AI systems identify situations where humans are likely to be unable to accurately compute the ``complete picture''? Could this effect be leveraged by AI algorithms to learn effectively from smaller datasets?  \\ 

\midrule

%------------------------(Interpretation)------------------------

\parbox[t]{2mm}{\multirow{4}[45]{*}{\rotatebox[origin=c]{90}{\small\textbf{Interpretation}}}}

& \emph{Conjunction} fallacy \cite{Tversky1983,Tentori2004,Wedell2008,Lo2002} & In certain situations, humans see the conjunction of two events as being more likely than any one event individually & Could AI systems recognize situations where humans are likely to make such errors and provide alternate decisions?  \\ \cmidrule{2-4}
 
& \emph{Base Rate} fallacy \cite{Barbey2007,BarHillel1980} & Humans have a tendency to ignore the base rate information when making decisions & Human reasoning does not follow Bayesian reasoning in certain settings. Could AI systems leverage these non-Bayesian computations effectively? \\ \cmidrule{2-4}

& \emph{Gamblers} fallacy \cite{Tversky1974,Gold1997,Barron2010,Chen2016} & Humans tend to overvalue the impact of past events when predicting the outcome of independent future events & Decision making systems that learn from human decision making --e.g. legal, college admissions or HR decision-making systems-- learn from data that reflects the gamblers fallacy. How could this bias be mitigated to design fairer AI-based decision-support systems? \\ \cmidrule{2-4}

& \emph{Hyperbolic discounting} effect \cite{Thaler1981,Hampton2018,Ainslie1975} & Humans tend to choose immediate rewards over rewards that come later in the future & Studies have shown a link between high social media usage and hyperbolic discounting leading to unhealthy behavior \cite{Kurz2022,SchulzvanEndert2022}. Could AI systems recognize when we are impacted by this bias and help mitigate it?\\

\bottomrule

\end{tabular}

}
\caption{Selected biases in \textit{Presentation} and \textit{Interpretation} categories of the proposed taxonomy and their relevance to the study of AI systems.}
\label{table:biasesTablePart1}
\end{table}

\begin{table}[t]
\centering
\resizebox{\linewidth}{!}{

\begin{tabular}{ l|p{0.13\linewidth}|p{0.37\linewidth}|p{0.50\linewidth}}

\toprule

 & \multicolumn{1}{c|}{\small\textbf{Bias}} & \multicolumn{1}{c|}{\small\textbf{Brief Description}} & \multicolumn{1}{c}{\small\textbf{Relevance to AI}}\\

\midrule

%------------------------(Value Attribution)------------------------

\parbox[t]{2mm}{\multirow{4}[40]{*}{\rotatebox[origin=c]{90}{\small\textbf{Value Attribution}}}}

& \emph{Halo} effect \cite{Dion1972,Nisbett1977,Gibson2016,Landy1974} & Positive attributes associated with a person in one setting carry over other settings & Could this effect be utilized to create systems that are easier to trust? Does the halo effect manifest itself when humans interact with chatbots or robots? \\ \cmidrule{2-4}
 
& \emph{IKEA} effect \cite{Norton2012, Radtke2019, Brunner2022} & Humans associate a higher value to their own creations than those of others  & Could this effect be leveraged to provide explanations that users are more likely to accept? \\ \cmidrule{2-4}
 
& \emph{Risk aversion} bias \cite{Pratt1978,Stanovich2010,Fischhoff1978,Rottenstreich2001} & We tend to avoid risky decisions even if they have a higher net expected utility than less risky options, especially if the uncertainty is high & Could AI systems support human decision-making by counter-balancing the risk aversion risk? \\ \cmidrule{2-4}
 
& \emph{Social desirability} bias \cite{Crowne1960,Herbert1995,Harrison2006,Stuart2009} & Humans tend to provide the answers to surveys or questions that they believe are expected from them & Do people provide socially desirable answers even when they are interacting with or being evaluated by machines? If yes, could the social desirability bias be be leveraged to nudge users to improve their behavior? \\ %\cmidrule{2-4}

\midrule

%------------------------(Recall)------------------------

\parbox[t]{2mm}{\multirow{4}[35]{*}{\rotatebox[origin=c]{90}{\small\textbf{Recall}}}}

& \emph{False memory} bias \cite{Loftus1974,Loftus1975,Lampinen1997} & Humans  incorrectly remember a past event depending on the questions they are asked about the event & False memories impact how we make decisions. Positive false memories have been shown to have positive consequences \cite{Bernstein2009}. Could this AI systems use this effect to improve user experience? \\ \cmidrule{2-4}

& \emph{Self-reference} effect \cite{Rogers1977,Gutchess2007} & Events with a direct impact are more likely to be remembered & Could AI systems leverage this effect to make explanations about their behavior more ``memorable''? \\ \cmidrule{2-4}

& \emph{Serial-positioning} effect \cite{Murdock1962,Murdock1978,Asch1946} & Items at the start and the end of a list are more memorable than those in the middle & When providing explanations, could AI systems leverage this bias to have a more effective human interaction?  \\ \cmidrule{2-4}

& \emph{Peak-end} rule \cite{Kahneman1993,Carmon1996,DeMaeyer2013} & The value of an event tends to be assessed based on its peak and final values, neglecting other parameters, such as its duration or average value & Could this bias be leveraged by AI systems to increase the perceived utility of hard tasks? Instead of maximizing the time a user stays on a social media platform, could this bias be used to reduce time spent while also increasing user satisfaction? \\ 

\bottomrule

\end{tabular}

}
\caption{Selected biases in the \textit{Value Attribution} and \textit{Recall} categories of the proposed taxonomy and their relevance to the study of AI systems.}
\label{table:biasesTablePart2}
\end{table}

\begin{table}[t]
\centering
\resizebox{\linewidth}{!}{

\begin{tabular}{ l|p{0.13\linewidth}|p{0.37\linewidth}|p{0.50\linewidth}}

\toprule

 & \multicolumn{1}{c|}{\small\textbf{Bias}} & \multicolumn{1}{c|}{\small\textbf{Brief Description}} & \multicolumn{1}{c}{\small\textbf{Relevance to AI}}\\

\midrule

%------------------------(Decision)------------------------

\parbox[t]{2mm}{\multirow{5}[40]{*}{\rotatebox[origin=c]{90}{\small\textbf{Decision}}}}

& \emph{Status quo} bias \cite{Samuelson1988,Kahneman1991} & Humans tend to make decisions that maintain the current state rather than changing it & Are AI systems that recommend fewer changes more likely to be trusted? \\ \cmidrule{2-4}

& \emph{Shared information} bias \cite{Forsyth1990,Postmes2001,Stasser1992} & In group settings, humans tend to focus the discussion on information everyone already has rather than trying to bring in new information & Could AI systems leverage this bias to effectively drive group conversations? \\ \cmidrule{2-4}
 
& \emph{Naive} allocation \cite{Simonson1990,Read1995,Kliger2014} & People tend to allocate resources equally between all options rather than based on the value of the options & Could AI systems provide alternatives to avoid naive allocation? Alternatively, could AI systems leverage this bias effectively to make decisions in high uncertainty situations? \\ \cmidrule{2-4}
 
& \emph{Take-the-best} heuristic \cite{Gigerenzer1996} & The decision between two alternatives is made based on the first cue that discriminates them & Could this heuristic be implemented in AI systems to learn efficiently from small datasets? \cite{Wang2022} \\ 

\bottomrule

\end{tabular}

}
\caption{Selected biases in the \textit{Decision} category of the proposed taxonomy and their relevance to the study of AI systems.}
\label{table:biasesTablePart3}
\end{table}

\section{Cognitive Biases and AI: Research Directions}
\label{sec:rqs}

Given the ubiquity of AI-based systems in our daily lives --from recommender systems to personal assistants and chatbots-- and the pervasiveness of our cognitive biases, there is an opportunity to leverage cognitive biases to build more efficient AI systems. 

In this section, we propose three research directions to further explore the interplay between cognitive biases and AI: (1) Human-AI interaction, (2) Cognitive biases in AI algorithms and (3) Computational modeling of cognitive biases.

\subsection{Area I. Human-AI Interaction}
\label{sec:rqs.area1}

Cognitive biases have been studied since the 1970's in experiments where human participants interacted with other humans, animals or inanimate objects. However, as Hidalgo et al.~\cite{Hidalgo2021} note, we do not necessarily perceive, interact with and evaluate machines in the same way as we do with humans, animals or objects. Thus, it is unclear today whether these cognitive biases exist when humans interact with AI systems, and if so with which degree of intensity and under what circumstances. 

This is especially the case with biases related to presentation and decision-making, as per Figure \ref{fig:biasesTaxonomy}. Previous work has reported that humans are influenced by observing machine behavior. For example, Hang, Ono, and Yamada~\cite{Hang2021} showed that participants who saw a video of robots exhibiting altruistic behavior were more likely to demonstrate altruistic behavior themselves. Others suggest that humans regard machines as social entities if they display ``sufficient interactive and social cues'' \cite{Dou2021}; and a third set of studies propose that humans view machines as being different from themselves in social interactions \cite{Straten2020}. Given the impact that cognitive biases have in many of our daily tasks and given the increased presence of AI algorithms to tackle many of these tasks, it becomes important to understand whether interactions with AI systems exhibit the same biases as those observed in human-to-human interactions.

For example, according to the \emph{social desirability bias}, users tend to respond to surveys with socially expected answers which are not necessarily their honest responses. Would this bias also emerge when users interact with a chatbot? Another suitable bias to study in this context is the \emph{halo} effect. Does it exist when humans interact with chatbots, robots or avatars? Are positive traits associated with an AI system in one area carried forward to other areas as well or are machines viewed as tools with a single purpose and hence outside of the scope to the halo affect? 

In addition to verifying if cognitive biases exist in human-machine interactions, these biases could be leveraged to design more human-like AI systems. While it has been reported that humans rely on heuristics when deciding if they should trust AI decisions \cite{Lu2021}, would the existence of cognitive biases in AI systems have an impact on their interpretability and trustworthiness? Anthropomorphic agents have been shown to increase user satisfaction and increase acceptance \cite{Pizzi2021}. However, studies on anthropomorphism tend to focus on physical attributes. We propose exploring anthropomorphism in the realm of cognitive biases. Exemplary biases that could be studied include the \emph{framing} effect, which could inform the fine tuning of language models in chatbots; the \emph{status quo} bias, which could increase the trust in AI systems that suggest small rather than major changes; or the \emph{halo} effect, which could lead to humans trusting chatbots, avatars or robots with certain appearances and attributes more than others independently of their actual performance.

Another dimension worth exploring is the intersection between cognitive biases and the explainability of AI systems (XAI). While there is a large body of previous work in XAI, only a small subset of cognitive biases has been considered in this research area \cite{Wang2019, Miller2019}, mainly, with the objective of mitigating them. Bu{\c{c}}inca, Malaya, and Gajos~\cite{Bucinca2021} note that many machine explanations are not useful because users rely on heuristics about when to trust the machine and when not to, rather than using the provided explanation to make such a decision. The authors propose cognitive forcing functions to help users consider machine explanations carefully and show the effectiveness of these functions in certain scenarios through user studies. 
We propose expanding the research agenda to consider the inclusion of cognitive biases in XAI. Exemplary biases in this context are the \emph{framing effect} and the \emph{self-reference effect} which AI models could potentially leverage to provide more trustworthy explanations.

Finally, we postulate that it could be valuable to include knowledge about human cognitive biases when designing AI systems that interact with users. Examples of biases that could be considered by AI systems include the \emph{gambler's fallacy}, the \emph{anchoring} effect or the \emph{framing} effect, by presenting information to users in a manner that would mitigate the existence of these biases; the \emph{default} heuristic, by nudging users to consider all options or highlighting the existence of alternative possibilities; and the \emph{shared information} bias, by performing a topic analysis on human conversations and providing hints on novel topics to be discussed.

\subsection{Area II. Cognitive Biases in AI Algorithms}
\label{sec:rqs.area2}

While cognitive biases and heuristics lead to sub-optimal decision making in certain situations, they are undoubtedly useful decision making aids. These heuristics are at times as effective as complex decision making rules while at the same time significantly reduce our cognitive load \cite{gg_hh}. Given that humans benefit greatly from the use of cognitive biases and heuristics, it is worth exploring how AI could also benefit from them.

Taniguchi, Sato, and Shirakaw~\cite{Taniguchi2017} work in this direction by building a modified Naive Bayes classifier that leverages the symmetry \cite{Sidman1982} and the mutual exclusion \cite{Merriman1989} biases. The proposed model is able to perform better than alternative, state-of-the-art methods on a spam classification task when the dataset is small and biased. Taniguchi, Sato, and Shirakawa~\cite{Taniguchi2019} and Manome et al.~\cite{Manome2021} extend this idea by incorporating the same biases in neural networks and learning vector quantization respectively for different tasks.

Given these successful examples of incorporating two cognitive biases in the design of AI systems, it is worth considering how other biases could be leveraged to help design AI systems that would learn faster and from less data. Additional biases --beyond the symmetry and mutual exclusion biases-- that could be relevant for this purpose include the \emph{take-the-best} heuristic; \emph{naive allocation}; the \emph{status-quo} bias which humans use to make effective decisions in situations with a high degree of uncertainty; and the \emph{fast-and-frugal} heuristics \cite{Gigerenzer1999} which have been shown to be effective decision making tools in real world scenarios such as medical decision making \cite{Gigerenzer2005}. They could potentially be used in the design of AI systems as noted in their recent work \cite{Wang2022}. 

%TC:ignore
\begin{table*}[ht]
\centering
\begin{tabular}{|m{0.12\textwidth}|m{0.68\textwidth}|}
\hline
\textbf{Cognitive Bias} & \texttt{Confirmation Bias} \\
\hline
\textbf{Definition} & The tendency of individuals to seek, interpret, and remember information in a way that confirms their preexisting beliefs or hypotheses, while ignoring or dismissing contradictory evidence. \\
\hline
\textbf{Example} & Sam believes that a certain dietary supplement has remarkable health benefits. They regularly read online forums and articles that promote the supplement's positive effects, and they tend to ignore or dismiss any information suggesting potential risks or ineffectiveness. \\
\hline
\textbf{AI's Role} & 
\emph{Detecting the bias}: Through the collection and analysis of both relevant online content and Sam's information consumption behavior related to the dietary supplement, machine learning algorithms could be used to detect the existence of the confirmation bias. \\
& 
\emph{Counteracting the bias}: Machine learning-based recommendation algorithms could recommend diverse perspectives and evidence-based information to challenge Sam's preconceived notions and encourage a more balanced understanding of the supplement's properties. \\
\hline
\end{tabular}
\caption{Example of how AI could support humans in mitigating the confirmation bias.} \label{tab:conf}
\end{table*}
%TC:endignore

\subsection{Area III. Computational Modeling of Cognitive Biases}
\label{sec:rqs.area3}

The third research area addresses the computational modeling of cognitive biases.  Hiatt et al.~\cite{Hiatt2017} present a detailed survey of the computational approaches proposed to date to model human behavior in human-machine interaction. While they highlight the importance of having models that can account for ``some basic understanding of human reasoning, fallacy and error'', none of the approaches they present explicitly models cognitive biases, which we believe is crucial for the design of AI systems going forward.

Interestingly, in the past decade we have witnessed an exponential growth of research on understanding and mitigating algorithmic biases \cite{biases_in_fairness_1,biases_in_fairness_2} which are different from human cognitive biases. However, a failure to recognize the differences has led to misrepresentations \cite{Sen2020, Harris2020}.

Previous work has focused on building models of the underlying cognitive process that explain cognitive biases --such as bounded rationality \cite{Munier1999,Leibfried2016}. Alternatively, scholars have focused on a subset of biases and have proposed a variety of computational models in a particular task.
One of the most promising modeling frameworks in this context is Bayesian modeling \cite{natural_stupidity,Tenenbaum1998,Rastogi2022,Griffiths2006,Chater2006,Jansen2021}. Beyond Bayesian modeling, Kang and Lerman~\cite{vip_lerman} build on existing generative models to predict the relevance of an item while accounting for the position bias \cite{Blunch1984}.

Additional work has been proposed in the medical field: Crowley et al.~\cite{Crowley2012} define a set of $8$ biases as a sequence of events in computer-based pathology tasks.  McShane, Nirenburg, and Jarrel~\cite{McShane2013} provide a tool to support doctors in their diagnoses while mitigating recall biases. Alternative approaches are based on expert observations of subjects performing certain tasks \cite{Albert2016,Albert2018}. However, such methods are difficult to scale.

Proposing a unifying, task-independent AI-based framework to automatically identify cognitive biases from observed human behavior could have a profound impact in the design of AI systems. Such a framework could provide a representation that makes it possible to mitigate cognitive biases effectively. It could enable the development of personalized systems that support each individual by providing the most suitable mitigation strategy for them. 

In the persuasive computing literature, it has been observed that different people respond to different techniques to support them in modifying their behavior, from simple awareness to social support or competition \cite{Michie2017,deOliveira2010}. Recent work by Kliegr, Bahn{\'{\i}}k, and F\"{u}rnkranz~\cite{Kliegr2021} has studied 20 cognitive biases that could potentially impact how humans interpret machine learning models and proposes several debiasing techniques. While informing users about their biases is certainly a useful first step, it is generally not enough to mitigate them \cite{Heuer1999}. Personalization and persuasive computing methods could open new doors to more effective cognitive bias mitigation strategies.

\section{Conclusion}

Human perception, memory and decision-making are impact by cognitive biases and heuristics that influence our actions and decisions. Despite the pervasiveness of such biases, they are largely disregarded by today’s AI systems that model human behavior and interact with humans. However, given the increased prominence of AI-human collaboration, we believe that it would be crucial for AI systems to consider this fundamental element of human cognition. 

In this theoretical paper, we have proposed a taxonomy of cognitive biases from the perspective of AI-human collaboration and have selected four exemplary biases in each of the five key dimensions of the proposed taxonomy. We have also proposed three broad research areas in the intersection between AI and cognitive biases: First,  human-AI interaction which focuses on open questions, such as determining if human-AI interaction exhibits the same cognitive biases as human-to-human interaction and exploring the potential value of including cognitive biases in AI systems to make them more trustworthy and interpretable. Second, cognitive biases in AI systems, focused on the potential of leveraging the mechanisms behind our cognitive biases and heuristics to build more robust and efficient machine learning algorithms. Third, the computational modeling of cognitive biases to achieve a unifying modeling framework which could be used to design personalized mitigation strategies to support human decision making.

\section*{Acknowledgements}
Aditya Gulati and Nuria Oliver are supported by a nominal grant received at the ELLIS Unit Alicante Foundation from the Regional Government of Valencia in Spain (Convenio Singular signed with Generalitat Valenciana, Conselleria de Innovacion, Industria, Comercio y Turismo, Direccion General de Innovacion). Aditya Gulati is also supported by a grant by the Banc Sabadell Foundation.

\bibliography{references}

\end{document}